\documentclass[aps,prd,%multicol,twocolumn, 
preprintnumbers,groupedaddress]{revtex4}
\usepackage{epsbox}
\usepackage{amsmath}
\usepackage[dvips]{graphicx}

\begin{document}
\preprint{\vtop{
{\hbox{YITP-10-69}\vskip-0pt
                 \hbox{KANAZAWA-10-06} \vskip-0pt
%                 \hbox{hep-ph/1008.2992} 
}
}
}

%\date{\today}

\title{ 
Tetra-quark mesons with exotic quantum numbers}

\author{
Kunihiko Terasaki   %authors' name%
}
\affiliation{
Yukawa Institute for Theoretical Physics, Kyoto University,
Kyoto 606-8502, Japan \\
Institute for Theoretical Physics, Kanazawa University, 
Kanazawa 920-1192, Japan
}

\begin{abstract}
{
Tetra-quark mesons with exotic quantum numbers, their production 
rates and decay properties are studied, because they are useful to 
establish existence of tetra-quark mesons. 
}
\end{abstract}

\maketitle

Tetra-quark mesons can be classified into the following four groups 
in accordance with difference of symmetry property of their flavor  
wavefunctions (wfs.)~\cite{Jaffe} 
 (and \cite{D_{s0}-KT}), 
%%%%%%%%%%%%%%%%%%%%%%%%%%%%%%%%%%%%%%%%%%%%%%%%%%%%%%%%%%%%%%%%%%%%%%%%
\begin{eqnarray} 
&&\hspace{-8mm} \{qq\bar q\bar q\} =  
[qq][\bar q\bar q] \oplus (qq)(\bar q\bar q)  
\oplus \{[qq](\bar q\bar q)\oplus (qq)[\bar q\bar q]\} 
                                                   \label{eq:4-quark} 
\end{eqnarray} 
%%%%%%%%%%%%%%%%%%%%%%%%%%%%%%%%%%%%%%%%%%%%%%%%%%%%%%%%%%%%%%%%%%%%%%%%
with $q=u,d,s$ (and $c$), where parentheses and square brackets denote 
symmetry and anti-symmetry, respectively, of flavor wfs. under exchange 
of flavors between them. 
Each term on the right-hand-side (r.h.s.) of Eq.~(\ref{eq:4-quark}) is 
again classified into two groups with~\cite{Jaffe}  
%%%%%%%%%%%%%%%%%%%%%%%%%%%%%%%%%%%%%%%%%%%%%%%%%%%%%%%%%%%%%%%%%%%%%%%%
${\bf \bar 3_c}\times{\bf 3_c}$ and ${\bf 6_c}\times {\bf \bar 6_c}$  
%%%%%%%%%%%%%%%%%%%%%%%%%%%%%%%%%%%%%%%%%%%%%%%%%%%%%%%%%%%%%%%%%%%%%%%%
of the color $SU_c(3)$, because these two can lead to color-singlet 
tetra-quark states. 
The force between two quarks $qq$ is attractive (or 
repulsive)~\cite{color} when they are of $\bf{\bar 3_c}$ (or $\bf{6_c}$) 
of $SU_c(3)$, so that the ${\bf \bar 3_c}\times{\bf 3_c}$ state is taken 
as the (dominant part of) lower lying one, unless these states largely 
mix with each other~\cite{D_{s0}-KT}. 
Spin ($J$) and parity ($P$) of (dominant components of) 
$[qq][\bar q\bar q]$ and $[qq](\bar q\bar q)\pm (qq)[\bar q\bar q]$ 
mesons are $J^P=0^+$ and $1^+$, respectively, while $(qq)(\bar q\bar q)$ 
can have~\cite{Jaffe,X-3872-KT} $J^P = 0^+,\,1^+,\,2^+$. 
However, we ignore this $(qq)(\bar q\bar q)$ because no signal of scalar 
$(K\pi)_{I=3/2}$ state which  can come from the $(qq)(\bar q\bar q)$ 
state~\cite{Jaffe} has been observed in the region $\lesssim 1.8$ 
GeV~\cite{K-pi-3/2}, although there is an argument that an iso-tensor 
state might have been observed around 1.6 GeV in the 
$\gamma\gamma^*\rightarrow \rho\rho$~\cite{LEP}. 

In inclusive $e^+e^-$ 
annihilation~\cite{Babar-e^+e^-D_{s0},CLEO-D_{s0}}, the charm-strange 
scalar $D_{s0}^+(2317)$ was discovered in the $D_s^+\pi^0$ mass 
distribution, while no signal has been observed in the $D_s^{*+}\gamma$ 
channel, and, therefore, the above results lead to a severe 
constraint~\cite{CLEO-D_{s0}},  
%%%%%%%%%%%%%%%%%%%%%%%%%%%%%%%%%%%%%%%%%%%%%%%%%%%%%%%%%%%%%%%%%%%%%%%%
\begin{eqnarray}
&&   \frac{\Gamma(D_{s0}^+(2317) \rightarrow D_{s}^{*+}\gamma)}
{\Gamma(D_{s0}^+(2317) \rightarrow D_{s}^{+}\pi^0)}\Biggl|_{\rm CLEO}
< 0.059.     
                                             \label{eq:ratio-D_{s0}}
\end{eqnarray}
%%%%%%%%%%%%%%%%%%%%%%%%%%%%%%%%%%%%%%%%%%%%%%%%%%%%%%%%%%%%%%%%%%%%%%%%
This implies~\cite{HT-isospin} that $D_{s0}^+(2317)$ is an iso-triplet 
state. 
The simplest way~\cite{D_{s0}-KT} to realize $D_{s0}^+(2317)$ as an 
iso-triplet scalar state is to assign it to 
$[cn][\bar s\bar n]_{I=1},\,(n=u,\,d)$ in Eq.~(\ref{eq:4-quark}) which 
is denoted as $\hat F_I^+$. 
In this case, the observed narrow width of $D_{s0}^+(2317)$ 
is understood by a small overlap of color and spin 
wfs.~\cite{KT-Hadron-2003,HT-isospin,ECT-talk}.  
The above assignment is consistent~\cite{QNP06-KT,Scadron-70,NFQCD-KT} 
with the observations in $B$ decays, i.e., signals of charm-strange 
scalar mesons have been observed in the $D_s^{*+}\gamma$ channel as 
well as in the $D_s^{+}\pi^0$, and branching fractions of $B$ 
decays producing them have been measured as~\cite{Belle-D_{s0}} 
%%%%%%%%%%%%%%%%%%%%%%%%%%%%%%%%%%%%%%%%%%%%%%%%%%%%%%%%%%%%%%%%%%%%%%%%
$Br(B\rightarrow\bar{{D}}
\tilde{{D}}_{{s0}}^{{+}}{(2317)}[{D_s}^{{+}}{\pi^0}]) = 
(8.5^{{+}2.1}_{{-}1.9} {\pm} 2.6)\times 10^{-4}$ and 
${Br}({B}\rightarrow \bar{{D}}\tilde{{D}}_{{s0}}^{{+}}{(2317)}
{[D_s^{{*+}}{\gamma}])}                                  
{\,\,=(2.5^{{+}2.0}_{{-}1.8}({<} 7.5))\times 10^{-4}}$, 
%%%%%%%%%%%%%%%%%%%%%%%%%%%%%%%%%%%%%%%%%%%%%%%%%%%%%%%%%%%%%%%%%%%%%%%%
where these signals observed in the $D_s^{*+}\gamma$ and $D_s^+\pi^0$ 
are denoted as 
%%%%%%%%%%%%%%%%%%%%%%%%%%%%%%%%%%%%%%%%%%%%%%%%%%%%%%%%%%%%%%%%%%%%%%%%
$\tilde{D}_{s0}^{+}(2317)[D_s^{*+}\gamma]$ and 
$\tilde{D}_{s0}^{+}(2317)[D_s^{+}\pi^0]$, 
%%%%%%%%%%%%%%%%%%%%%%%%%%%%%%%%%%%%%%%%%%%%%%%%%%%%%%%%%%%%%%%%%%%%%%%%
respectively.   
Therefore, it is natural to identify $\hat F_I^+$ and its iso-singlet 
partner $\hat F_0^+\sim [cn][\bar s\bar n]_{I=0}$ with 
$\tilde{D}_{s0}^{+}(2317)[D_s^{+}\pi^0]$ and  
$\tilde{D}_{s0}^{+}(2317)[D_s^{*+}\gamma]$, respectively, because   
$\hat F_I^+$ decays dominantly into $D_s^+\pi^0$ while $\hat F_0^+$ into 
$D_s^{*+}\gamma$. 
For more details, see Refs.~\cite{D_{s0}-KT,HT-isospin,ECT-talk} and 
\cite{NFQCD-KT}. 

Another candidate~\cite{Terasaki-X,omega-rho-KT} of tetra-quark meson 
is $X(3872)$ with~\cite{Belle-X-J^P} $J^P=1^+$, as seen below. 
It was discovered in the $\pi^+\pi^-J/\psi$ mass distribution by the 
Belle~\cite{Belle-X(3872)}, and confirmed~\cite{confirm-X} by the CDF, 
D0 and Babar. 
(Hereafter, we describe $J/\psi$ as $\psi$ to save space.) 
Its charge conjugation parity ($\mathcal{C}$) can be even, because it 
decays into the 
$\gamma\psi$~\cite{Belle-X-gamma-psi,Babar-X-gamma-psi}. 
However, it decays into two states with opposite 
$G$-parities\cite{Belle-X-gamma-psi}, 
%%%%%%%%%%%%%%%%%%%%%%%%%%%%%%%%%%%%%%%%%%%%%%%%%%%%%%%%%%%%%%%%%%%%%%%%
\begin{equation}
%R\equiv 
\frac{Br(X(3872)\rightarrow \pi^+\pi^-\pi^0\psi)}
{Br(X(3872)\rightarrow \pi^+\pi^-\psi)}
=1.0\pm 0.4\pm 0.3.                           \label{eq:3pi/2pi-exp}
\end{equation}
%%%%%%%%%%%%%%%%%%%%%%%%%%%%%%%%%%%%%%%%%%%%%%%%%%%%%%%%%%%%%%%%%%%%%%%%
This is puzzling because the well-known strong interactions conserve 
$G$-parity. 
In addition, it has been noted~\cite{Belle-X(3872),CDF-pipi} that 
%the Belle~\cite{Belle-X(3872)} and CDF~\cite{CDF-pipi} have noted that 
the decay $X(3872)\rightarrow \pi^+\pi^-\psi$ proceeds through 
$\rho^0\psi$. 
Because a search for its charged partners $X(3872)^\pm$ has given a 
negative result~\cite{Babar-X-charged-partner}, however, it would be 
an iso-singlet state, and isospin conservation does not work in the  
decay. 
Besides, it has been pointed out~\cite{Belle-X-gamma-psi,Babar-X-omega} 
that the $X(3872)\rightarrow\pi^+\pi^-\pi^0\psi$ decay proceeds 
through the sub-threshold 
$X(3872)\rightarrow \langle\omega \psi\rangle$. 
If isospin is conserved in this decay, $X(3872)$ would be an 
iso-singlet state. 
This is consistent with the above negative result on %the search for 
$X(3872)^\pm$. 

Although various approaches~\cite{various-approaches,Maiani} have been 
proposed to solve the above puzzle, they are unnatural because the 
well-known $\omega\rho^0$ mixing~\cite{omega-rho} as the origin of the 
isospin non-conservation in nuclear forces, which is 
compatible~\cite{omega-rho-KT} with the measured rate for the 
$\omega\rightarrow \pi^+\pi^-$ decay~\cite{PDG08}, has not been 
considered. 
Assuming that $X(3872)$ is a tetra-quark system 
like $\{[cn](\bar c\bar n) + (cn)[\bar c\bar n]\}_{I=0}$ 
meson~\cite{Terasaki-X} (or a $D^0\bar D^{*0}$ molecule) 
with even 
$\mathcal{C}$-parity and that the isospin non-conservation under 
consideration is caused by the 
$\omega\rho^0$ mixing~\cite{omega-rho-KT},     %i.e., 
%%%%%%%%%%%%%%%%%%%%%%%%%%%%%%%%%%%%%%%%%%%%%%%%%%%%%%%%%%%%%%%%%%%%%%%%
%$X(3872)\rightarrow \langle\omega\psi\rangle \rightarrow \rho^0\psi
%\rightarrow \pi^+\pi^-\psi$, 
%%%%%%%%%%%%%%%%%%%%%%%%%%%%%%%%%%%%%%%%%%%%%%%%%%%%%%%%%%%%%%%%%%%%%%%%
we have reproduced~\cite{omega-rho-KT,NFQCD-KT} the measured ratios,   
%%%%%%%%%%%%%%%%%%%%%%%%%%%%%%%%%%%%%%%%%%%%%%%%%%%%%%%%%%%%%%%%%%%%%%%%
\begin{equation}
R^\gamma_{\rm Belle} = 0.14 \pm 0.05 \hspace{2mm}{\rm and}\hspace{2mm}
R^\gamma_{\rm Babar} = 0.33 \pm 0.12,  
                                    \label{eq:radiative-fraction-exp}
\end{equation}
%%%%%%%%%%%%%%%%%%%%%%%%%%%%%%%%%%%%%%%%%%%%%%%%%%%%%%%%%%%%%%%%%%%%%%%%
given by the Belle~\cite{Belle-X-gamma-psi} and 
Babar~\cite{Babar-gamma-psi'}, respectively, where $R^\gamma$ is 
defined by 
%%%%%%%%%%%%%%%%%%%%%%%%%%%%%%%%%%%%%%%%%%%%%%%%%%%%%%%%%%%%%%%%%%%%%%%%
$R^\gamma\equiv {Br(X(3872)\rightarrow \gamma\psi)}/  
{Br(X(3872)\rightarrow \pi^+\pi^-\psi)}$. 
%%%%%%%%%%%%%%%%%%%%%%%%%%%%%%%%%%%%%%%%%%%%%%%%%%%%%%%%%%%%%%%%%%%%%%%%
In the above, we have considered the $X(3872)\rightarrow \gamma\psi$ 
decay in place of $X(3872)\rightarrow \pi^+\pi^-\pi^0\psi$ in 
Eq.~(\ref{eq:3pi/2pi-exp}), because both of them are controlled by the 
same subthreshold $X(3872)\rightarrow \langle\omega\psi\rangle$ decay 
when the vector meson dominance (VMD)~\cite{VMD} is applied to the 
radiative decay, while the kinematics of the former is much simpler 
than the latter. 
In contrast, if $X(3872)$ were assumed to be a charmonium 
$X_{c\bar c}$, the $\psi$ pole contribution, 
$X_{c\bar c}\rightarrow\psi\psi\rightarrow\gamma\psi$, 
would be dominant in the radiative decay because of the OZI 
rule~\cite{OZI}, and as the result, 
%%%%%%%%%%%%%%%%%%%%%%%%%%%%%%%%%%%%%%%%%%%%%%%%%%%%%%%%%%%%%%%%%%%%%%%%
$(R^\gamma_X)_{{c\bar c}}\gg 
(R^\gamma_X)_{\rm Babar} \sim (R^\gamma_X)_{\rm Belle}$ 
%%%%%%%%%%%%%%%%%%%%%%%%%%%%%%%%%%%%%%%%%%%%%%%%%%%%%%%%%%%%%%%%%%%%%%%%
would be obtained~\cite{omega-rho-KT,NFQCD-KT}. 
Therefore, we see that a tetra-quark interpretation of $X(3872)$ is 
favored over the charmonium, although a small mixing of $X_{c\bar c}$ 
is not excluded. 
In addition, it should be noted that production~\cite{CDF-prompt-X} of 
the \underline{prompt} $X(3872)$ seems to favor a compact object like 
a tetra-quark meson over a loosely bound 
molecule~\cite{compact-prompt-X}.   
Although an argument against the above conclusion was 
proposed~\cite{Braaten-propmt-X},  
it does not explicitly prove that a loosely bound molecule provides a 
sufficiently large cross section for the prompt $X(3872)$ production. 
In addition, the $D^0\bar D^{*0}$ molecule involves some other 
problems~\cite{Terasaki-X}.  
%%%%%%%%%%%%%%%%%%%%%%%%%%%%%%%%%%%%%%%%%%%%%%%%%%%%%%%%%%%%

Tetra-quark interpretations of $D_{s0}^+(2317)$ and $X(3872)$ have been 
favored by experiments as seen above, although their quantum numbers 
are not exotic. 
Therefore, observation of their partners would be needed to establish 
the above interpretations. 
However, for example, neutral and doubly charged partners, $\hat F_I^0$ 
and $\hat F_I^{++}$, of $\hat F_I^+$ 
have not been observed in inclusive $e^+e^-$ 
annihilation~\cite{Babar-D_{s0}-charged-partners}. 
Nevertheless, it does not imply their non-existence but 
a \underline{suppression of their production} in this process.  
This can be understood by considering a production mechanism in the 
framework of minimal $q\bar q$ pair creation~\cite{Scadron-70}.  
On the other hand, in $B$ decays,  
their production rates have been estimated 
as~\cite{production-D_{s0}-KT,Scadron-70},  
%%%%%%%%%%%%%%%%%%%%%%%%%%%%%%%%%%%%%%%%%%%%%%%%%%%%%%%%%%%%%%%%%%%%%%%%
$Br(B_u^+\rightarrow D^-\hat F_I^{++})
\sim Br(B_d^0\rightarrow \bar D^0\hat F_I^{0})  
\sim Br(B_u^+(B_d^0)\rightarrow \bar D^0(D^-)\hat F_0^+)         
\sim Br(B_u^+(B_d^0)\rightarrow \bar D^0(D^-)\hat F_I^+)_{\rm exp}  
\sim 10^{-(4 - 3)}$,     
%%%%%%%%%%%%%%%%%%%%%%%%%%%%%%%%%%%%%%%%%%%%%%%%%%%%%%%%%%%%%%%%%%%%%%%%
because the above decays are described by the same type of quark-line 
diagrams and hence the sizes of their amplitudes are expected to be 
nearly equal to each other. 
The results are large enough to observe them. 

Another evidence for existence of tetra-quark mesons can be obtained by 
observing mesons with exotic quantum number(s). 
In the present scheme, $\hat E^0\sim [cs][\bar u\bar d]$ 
meson~\cite{D_{s0}-KT} is 
%where $(qq)(\bar q\bar q)$ mesons are ignored,  
only one scalar meson with an exotic set ${C} = -S = +1$ of 
quantum numbers.    % can exist.  
Axial-vector mesons with exotic quantum numbers, which come from 
$\{[qq](\bar q\bar q)\oplus (qq)[\bar q\bar q]\}$ in 
Eq.~(\ref{eq:4-quark}), are 
$H_{Acc}^+\sim (cc)[\bar u\bar d]$ with $C=2$, $S=0$, $I=0$; 
$K_{Acc}^{++}\sim (cc)[\bar d\bar s]$ and 
$K_{Acc}^+\sim(cc)[\bar u\bar s]$ with $C=2$, $S=1$, $I=1/2$; 
$E^0_{A(cs)}\sim(cs)[\bar u\bar d]$ with $C=1$, $S=-1$, $I=0$; 
$E_{A[cs]}^+\sim[cs](\bar d\bar d)$,  
$E_{A[cs]}^0\sim[cs](\bar u\bar d)$ and 
$E_{A[cs]}^-\sim[cs](\bar u\bar u)$ with $C=1$, $S=-1$, $I=1$. 
%%%%%%%%%%%%%%%%%%%%%%%%%%%%%%%%%%%%%%%%%%%%%%%%%%%%%%%%%%%%%%%%%%%%%%%%
Their masses can be very crudely estimated by using a quark counting 
with $m_c - m_s\simeq 1.0$ GeV and $m_s - m_n\simeq 0.1$ GeV, as in 
Ref.~\cite{D_{s0}-KT}. 
The results are listed in Table 1. 
It should be noted that the same quark counting has 
predicted~\cite{hidden-charm-scalar-KT} the mass of hidden-charm scalar 
$\hat\delta^c_{I=1}$ as $m_{\hat\delta^c_{I=1}}\simeq 3.3$ GeV. 
Our prediction matches well to a peak in the $\eta\pi$ channel at 
$3.2$ GeV observed by the Belle~\cite{hidden-charm-scalar-Belle}, 
although it is much lower than  predictions by the other 
models~\cite{hidden-charm-scalar,Maiani}. 
%%%%%%%%%%%%%%%%%%%%%%%%%%%%%%%%%%%%%%%%%%%%%%%%%%%%%%%%%%%%%%%%%%%%%%%%
\begin{figure}[!b]    %
%\begin{center}    %
\hspace{5mm}
\includegraphics[width=130mm,clip]{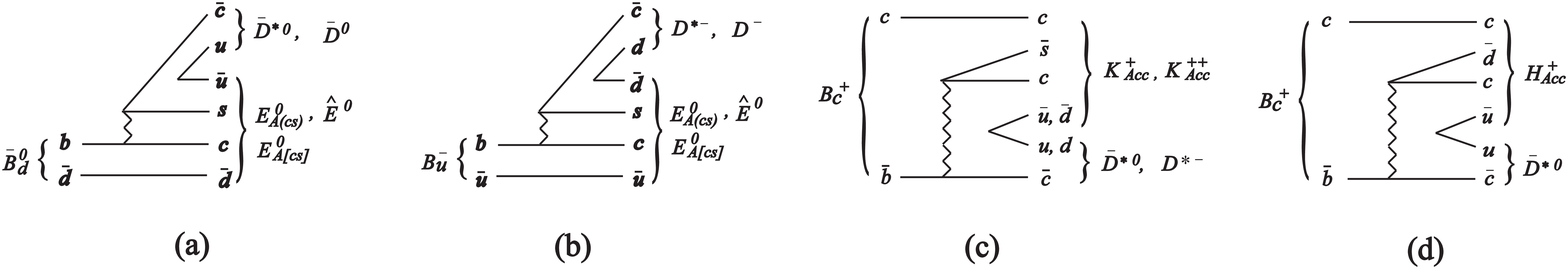}
\label{fig:prod-Exotic-ptp.eps}                
\begin{quote} 
Fig.~1. Productions of tetra-quark scalar $\hat E^0$ and axial-vector 
$E_{A(cs)}^0$, $E_{A[cs]}^{-,0,+}$, $K_{Acc}^{+(+)}$ and $H_{Acc}^+$ with 
exotic quantum numbers.    
\end{quote}    \vspace{-5mm}
\end{figure}%      
%%%%%%%%%%%%%%%%%%%%%%%%%%%%%%%%%%%%%%%%%%%%%%%%%%%%%%%%%%%%%%%%%%%%%%%%

Now we consider productions of tetra-quark mesons with exotic quantum 
number(s) in the same framework as the productions of $D_{s0}^+(2317)$ 
and its partners in Ref.~\cite{production-D_{s0}-KT}. 
Center-of-mass momenta ($p$'s) of the daughters in the $B$ decays 
producing tetra-quark mesons listed above are of the same order of 
magnitude (in the region $0.9\lesssim p \lesssim 2.0$ GeV). 
Therefore, if all the sizes of amplitudes are of the same order of 
magnitude, all the rates for the decays under consideration, which are 
expected to be dominantly of $S$-wave, would be of the same order of 
magnitude except for the CKM suppressed $H_{Acc}^+$ production. 
Full widths of $B_u$, $B_d$ and $B_c$ are approximately proportional to 
(parent mass)$^5$, so that the width of $B_c$ is larger by about a 
factor 2.4 than those of $B_u$ and $B_d$ (but within the same order 
of magnitude). 
Therefore, amplitudes (with the same order of magnitude) for decays 
under consideration lead to branching fractions of the same order of 
magnitude. 

Productions of $K_{Acc}^+$ and $K_{Acc}^{++}$ listed above can be 
described by the quark-line diagram, Fig.~1(c), which is of the same 
type as Fig.~2(a) and Fig.~3(b) in Ref.~\cite{production-D_{s0}-KT} 
describing 
%%%%%%%%%%%%%%%%%%%%%%%%%%%%%%%%%%%%%%%%%%%%%%%%%%%%%%%%%%%%%%%%%%%%%%%%
$B_u^+\rightarrow \bar D^0\hat F_I^+$ and 
$B_d^0\rightarrow D^-\hat F_I^+$,  
%%%%%%%%%%%%%%%%%%%%%%%%%%%%%%%%%%%%%%%%%%%%%%%%%%%%%%%%%%%%%%%%%%%%%%%%
respectively. 
Because 
%%%%%%%%%%%%%%%%%%%%%%%%%%%%%%%%%%%%%%%%%%%%%%%%%%%%%%%%%%%%%%%%%%%%%%%%
$Br(B_u^+(B_d^0)\rightarrow \bar D^{0}(D^-)D_{s0}^+(2317))_{\rm exp}
\sim 10^{-(4 - 3)}$    
%%%%%%%%%%%%%%%%%%%%%%%%%%%%%%%%%%%%%%%%%%%%%%%%%%%%%%%%%%%%%%%%%%%%%%%%
as mentioned before, production rates for $K_{Acc}^+$ and $K_{Acc}^{++}$ 
would be very crudely estimated as 
%%%%%%%%%%%%%%%%%%%%%%%%%%%%%%%%%%%%%%%%%%%%%%%%%%%%%%%%%%%%%%%%%%%%%%%%
\begin{equation}Br(B_c^+\rightarrow D^{*-}K_{Acc}^{++})\sim 
Br(B_c^+\rightarrow \bar D^{*0}K_{Acc}^+)\sim 10^{-(4 - 3)}.
                                                                   \label{eq:rates-for-K_A}
\end{equation}  
%%%%%%%%%%%%%%%%%%%%%%%%%%%%%%%%%%%%%%%%%%%%%%%%%%%%%%%%%%%%%%%%%%%%%%%%
Production of $H_{Acc}^+$ with $I=0$ is described by the diagram 
Fig.~1(d). 
However, it is now the CKM suppressed decay, so that the rate for 
$H_{Acc}^+$ production would be more suppressed by a factor 
$\sim 1/10 - 1/20$ than the above case, although it is described by 
the same type of diagram. 
Productions of scalar and axial-vector tetra-quark mesons with 
${C} = -S = 1$ can be described by the diagrams, (a) and (b) in Fig.~1. 
These diagrams are of the same type as that of Fig.~4(c) in 
Ref.~\cite{production-D_{s0}-KT} describing 
$\bar B_d^0\rightarrow K^-\hat F_I^+$ whose rate also has already 
been measured as~\cite{BELLE-D_{s0}-K^-}   
%%%%%%%%%%%%%%%%%%%%%%%%%%%%%%%%%%%%%%%%%%%%%%%%%%%%%%%%%%%%%%%%%%%%%%%%
${Br}(\bar B_d^0\rightarrow K^-D_{s0}^+(2317))\cdot
{Br}(D_{s0}^+(2317)\rightarrow D_s^+\pi^0)        
=(5.3^{+1.5}_{-1.3}\pm 0.7\pm 1.4)\times 10^{-5}$.     
%%%%%%%%%%%%%%%%%%%%%%%%%%%%%%%%%%%%%%%%%%%%%%%%%%%%%%%%%%%%%%%%%%%%%%%%
%with $D_{s0}^+(2317) = \hat F_I^+$. 
By taking $Br(D_{s0}^+(2317)\rightarrow D_s^+\pi^0)\simeq 1$ as expected 
in Eq.~(\ref{eq:ratio-D_{s0}}), it follows that 
$Br(\bar B_d^0\rightarrow K^-D_{s0}^+(2317))\sim 10^{-(5-4)}$. 
However, this is smaller by about one order of magnitude than 
$Br(\bar B_d^0\rightarrow D^-D_{s0}^+(2317))_{\rm exp}$. 
This would be because the former includes an $s\bar s$ creation while 
the latter an $n\bar n$ creation (but no $s\bar s$) as discussed in 
Ref.~\cite{production-D_{s0}-KT}. 
In fact, a phenomenological analysis and a recent lattice QCD suggest that 
$s\bar s$ component is much smaller compared with $u\bar u$ and 
$d\bar d$ components in the sea quarks of nucleon ($0.07 - 0.22$ in 
the former~\cite{Bernard} and most likely $0.05$ in the latter~\cite{Ohki}). 
Therefore, we expect that the diagrams (a) and (b) in Fig.~1 would not 
be suppressed compared with the other ones, because they now involve 
no $s\bar s$ creation, i.e., 
%%%%%%%%%%%%%%%%%%%%%%%%%%%%%%%%%%%%%%%%%%%%%%%%%%%%%%%%%%%%%%%%%%%%%%%%
\begin{eqnarray}
&&
Br( B_u^-\rightarrow D^{*-}E^0_{A(cs)})                 
\sim Br( B_u^-\rightarrow D^{*-}E^0_{A[cs]}) 
\sim Br( B_u^-\rightarrow D^-\hat E^0)       \nonumber\\
&&
\sim Br(\bar B_d^0\rightarrow \bar D^{*0}E^0_{A(cs)})
\sim Br(\bar B_d^0\rightarrow \bar D^{*0}E^0_{A[cs]})
\sim Br(\bar B_d^0\rightarrow \bar D^0\hat E^0) \nonumber\\
&&\sim 10\times 
Br(\bar B_d^0\rightarrow K^-D_{s0}^+(2317))\sim 10^{-(4-3)},  
                                                                          \label{eq:rates-for-E}
\end{eqnarray}
%%%%%%%%%%%%%%%%%%%%%%%%%%%%%%%%%%%%%%%%%%%%%%%%%%%%%%%%%%%%%%%%%%%%%%%%
because all these decays are dominantly of $S$-wave and the 
center-of-mass momenta ($p$'s) of the daughter particles are of the same 
order of magnitude as discussed above.  
Therefore, the rates estimated above would be large enough to observe 
these exotic mesons, except for $H_{Acc}^+$. 
%%%%%%%%%%%%%%%%%%%%%%%%%%%%%%%%%%%%%%%%%%%%%%%%%%%%%%%%%%%%%%%%%%%%%%%%
\begin{table}[!t]
\begin{center}
\begin{quote}
{Table~1. %Possible decays of s
Scalar and axial-vector tetra-quark mesons with exotic quantum numbers. 
%, where $m_{\hat E}\simeq m_{\hat F_I^+} < m_D + m_K$ is assumed. 
Mass values are estimated by a quark counting.  
}
\end{quote}
\vspace{0.3cm}

\begin{tabular}
{ l | c | c | c | c | c }
\hline
%\begin{tabular}{c}  \\\vspace{3mm}
Tetra-quark %State 
%\end{tabular}
 &  
%\begin{tabular}{l}
mass %\\
%(GeV)
%\end{tabular} 
&
\multicolumn{4}{c}{Possible decays}      \\
\cline{3-6}
meson & (GeV) & 2-body & Threshold & 3-body & radiative (or weak)  \\
& & & \vspace{-4mm}\\
\hline
$\hat E^0$ & 2.32 & $\langle D\bar K\rangle$ 
& 2.36  &    & 
$(\bar K\pi)\bar K$, $(\bar K\pi\pi)\bar K$, $\cdots$
\\
\hline
& & & \vspace{-4mm}\\
$H_{Acc}^+$ & 3.87
& $D{D^*}$ & 3.88 & $DD\pi$ & $DD\gamma$
\\
\hline  
& \vspace{-4mm}\\
{$K_{Acc}$} &  3.97
& $ {D}{D_s^{*+}},\,D^*D_s^+ $ & 3.98 & ${D}{D_s}\pi$ & $D{D_s}\gamma$
\\
\hline 
& & & & & \vspace{-4mm}\\
$E_{A(cs)}$, $E_{A[cs]}$  & 2.97 & $\bar K{D^*},\,(\bar K^*D )$ 
& 2.61 (2.86 ) & $\bar K{D}\pi$ & $\bar KD\gamma$
\\
\hline  
%$E_{A[cs]}$ & 2.97
% & $\bar K{D^*},\,(\bar K^*D)$ & 2.61 (2.86 )  & $\bar KD\pi$ 
%& $\bar KD\gamma$
%\\
%\hline 
\end{tabular} \vspace{-2mm}
\end{center}
\end{table} 
%%%%%%%%%%%%%%%%%%%%%%%%%%%%%%%%%%%%%%%%%%%%%%%%%%%%%%%%%%%%%%%%%%%%%%%%

Although informations of decay properties of these exotic mesons in 
addition to their production rates would be useful to search for them, 
rates for two body decays of axial-vector mesons would be very 
sensitive to their mass values because they are (very) close to 
thresholds of their two body decays, 
%except for the 
%$E_{A[cs]},\,E_{A(cs)}\rightarrow \bar KD^*,\,\bar K^*D$ decays, 
as seen in Table~1.  
Therefore, calculations of rates for these two body decays would be 
inevitably model dependent at the present stage. 
In addition, no experimental data as the input data is known, so that 
they are left as our future subjects, and their possible radiative 
decays in addition to two- and three-body decays are listed, because 
they would be useful to search for these mesons. 
In the case of the scalar $\hat E^0$, its estimated mass 
has been lower than the $D\bar K$ threshold. 
It implies that $\hat E^0$ could decay neither through the ordinary 
strong interactions nor through the electromagnetic ones, because it 
has an exotic set of quantum numbers. 
Thus, it would decay through weak interactions~\cite{D_{s0}-KT}, for 
example, 
%%%%%%%%%%%%%%%%%%%%%%%%%%%%%%%%%%%%%%%%%%%%%%%%%%%%%%%%%%%%%%%%%%%%%%%%
$\hat E^0\rightarrow \langle D\bar K\rangle
\rightarrow (\bar K\pi)\bar K,\,\,
(\bar K\pi\pi)\bar K,\,\,(\bar K\ell\nu_\ell)\bar K$, etc.   
%%%%%%%%%%%%%%%%%%%%%%%%%%%%%%%%%%%%%%%%%%%%%%%%%%%%%%%%%%%%%%%%%%%%%%%%
Regarding the axial-vector mesons, however, their crudely estimated 
mass values are (very) close to  thresholds of possible two-body decays, 
so that possible three-body and radiative decays are additionally listed. 
Radiative three-body decays might be important in search for axial-vector 
mesons with exotic quantum numbers because of decay properties of 
$D^*$ and $D_s^{*+}$. 
%%%%%%%%%%%%%%%%%%%%%%%%%%%%%%%%%%%%%%%%%%%%%%%%%%%%%%%%%%%%%%%%%%%%%%%%

In summary, we have studied scalar and axial-vector mesons with exotic 
quantum numbers. 
We have estimated their production rates by comparing quark-line diagrams 
describing their productions with those of $D_{s0}^+(2317)$, and have seen 
that they can be large enough to observe in $B$ decays. 
Their possible decay modes are also studied. 
However, estimate of their rates is left as a subject in future, because no 
experimental data as the input data has not been known at the present 
stage. 

In the present scheme, $X(3872)$ has been interpreted as a tetra-quark 
$\{[cn](\bar c\bar n) + (cn)[\bar c\bar n]\}_{I=0}$ 
meson~\cite{Terasaki-X} with even $\mathcal{C}$-parity. 
To confirm this, it is also awaited to observe its opposite 
$\mathcal{C}$-parity partner 
$\{[cn](\bar c\bar n) - (cn)[\bar c\bar n]\}_{I=0}$ in the 
$\psi\pi^0\pi^0$ channel~\cite{NFQCD-KT}. 
%\vspace{-5mm}
%%%%%%%%%%%%%%%%%%%%%%%%%%%%%%%%%%%%%%%%%%%%%%%%%%%%%%%%%%%%%%%%%%%%%%%
\section*{Acknowledgments}    %\vspace{-3mm}
The author would like to thank Yukawa Institute for Theoretical 
Physics (YITP), Kyoto University. 
This work was motivated by discussions during the workshop on 
{\it New Frontier of QCD 2010 (NFQCD2010)}, Jan. 18 -- Mar. 19, 2010, 
YITP, Kyoto University, Kyoto, Japan. He also would like to 
appreciate Professor S.~Takeda for informing a recent analysis in 
nucleon $\sigma$ term from lattice QCD. 
%%%%%%%%%%%%%%%%%%%%%%%%%%%%%%%%%%%%%%%%%%%%%%%%%%%%%%%%%%%%%%%%%%%%%%%

%%%%%%%%%%%%%%%%%%%%%%%%%%%%%%%%%

%\end{references}
%%%%%%%%%%%%%%%%%%%%%%%
%%%%%%%%%%%%%%%%%%%%%%%
\end{document}